\documentclass[preprint,12pt]{elsarticle}
\usepackage{amssymb}
\usepackage{amsmath}
\usepackage{xcolor}
\usepackage{lineno}

\usepackage{natbib}
\usepackage{caption}
\usepackage{graphicx}
\usepackage[percent]{overpic} 
\usepackage{subcaption}    
\usepackage[final]{hyperref}
\journal{Results in Optics}

\begin{document}

\begin{frontmatter}

\title{ {An algorithm for super-resolved reconstruction of single-photon emitter locations from \texorpdfstring{$g^{(2)}(0)$}{g(2)(0)} maps} } 

\author{Sonali Gupta}
\author{Amit Kumar}
\author{Vikas S Bhat}
\author[cor]{Sushil Mujumdar}
\fntext[cor]{mujumdar@tifr.res.in}
\affiliation{organization = {Tata Institute of Fundamental Research},
    addressline = {Dr. Homi Bhabha Road, Colaba},
    city = {Mumbai},
    postcode = {400005},
    state = {Maharashtra},
    country = {India}}

\begin{abstract}
Single-photon sources are vital for emerging quantum technologies. 
In particular, Nitrogen-vacancy (NV) centers in diamond are promising due to their room-temperature stability, long spin coherence, and compatibility 
with nanophotonic structures. A key challenge, however, is the reliable 
identification of isolated NV centers, since conventional confocal 
microscopy is diffraction-limited and cannot resolve emitter distributions 
within a focal spot. Besides, the associated intensity scanning is a time-expensive procedure. Here, we introduce  {a simulation-based} raster-scanned $g^{(2)}(0)$ mapping technique combined 
with an inversion-based reconstruction algorithm.  {Using synthetically generated photon-timestamp data for randomly distributed NV centers under a scanned Gaussian excitation profile}, by directly measuring local 
photon antibunching across the field of view, we extract the effective emitter 
number within each focal spot and reconstruct occupancy maps on a 
sub-focal-spot grid. This enables recovery of the number and spatial 
distribution of emitters within regions smaller than the confocal focal spot, 
thereby offering possibilities of going beyond the diffraction limit.  { Simulations} confirm robust 
reconstruction of NV-center distributions.  {The method reduces experimental effort in identifying isolated emitters by inferring the effective emitter number and sub-diffraction occupancy within each focal spot in a single scan, thereby avoiding unnecessary searches in regions that do not contain single emitters, as commonly encountered in intensity-based scanning approaches.} It offers valuable 
feedback for nanophotonic device fabrication, supporting more precise and 
scalable integration of NV-based quantum photonic technologies.
\end{abstract}

\begin{graphicalabstract}
\includegraphics[width = \linewidth]{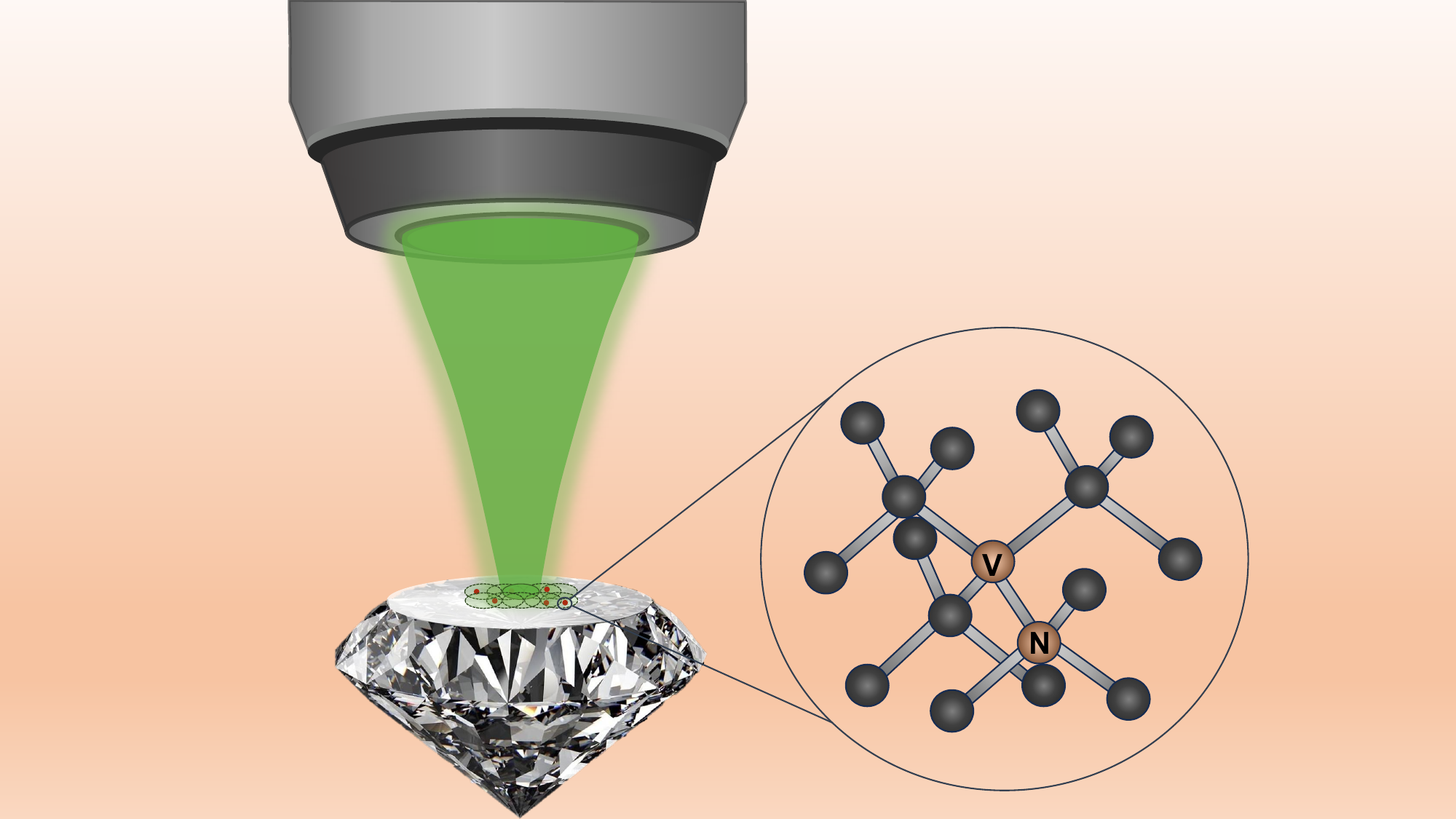}
\end{graphicalabstract}

\begin{highlights}
\item Reconstruction of the number and spatial distribution of emitters using raster-scanned $g^{(2)}(0)$ measurements and inversion-based algorithms.
\item Super-resolving emitter occupancy beyond the diffraction limit, enabling identification of emitter locations within a single diffraction-limited spot.
\item Quantitative validation of reconstruction accuracy under varying emitter densities and scan resolutions.
\end{highlights}

\end{frontmatter}

\section{Introduction}

Single-photon sources are indispensable for a broad range of quantum technologies, including quantum key distribution \cite{qcommunication}, quantum computation \cite{qcomputation}, quantum imaging \cite{qimaging}, quantum information processing \cite{qip}, and quantum metrology \cite{meterology}. A variety of physical platforms have been explored for on-demand single-photon generation, such as trapped ions \cite{PhysRevLett.58.203}, neutral atoms \cite{doi:10.1126/science.1113394}, molecules \cite{molecule}, semiconductor quantum dots \cite{quantumdot,qduppu}, two-dimensional materials like hexagonal boron nitride \cite{HBN}, and defect centers in solids \cite{SiV,NV}. Among these, Nitrogen Vacancy (NV) centers in diamond have emerged as particularly promising due to their stable fluorescence at room temperature \cite{roomtemp}, unique optical properties enabling spin initialization and single-shot readout \cite{initionalisation,singlreadout}, long spin-coherence times $T_2$ up to milliseconds at room temperature \cite{coherencetime} and up to one second at cryogenic temperatures \cite{1sec}, as well as magnetic-field sensitivity \cite{nvmagnetic,magneticimaging}. In addition, NV centers are readily integrated with existing nanofabricated devices and are compatible with photonic cavities \cite{Photonic,PC2}, waveguides \cite{waveguide}, and plasmonic structures \cite{Plasmonics}.

{Other solid-state single-photon emitter platforms like defect centers in hexagonal boron nitride (hBN) , emitters in transition metal dichalcogenides (TMDs) \cite{WSe2}, and semiconductor quantum dots also exhibit strong photon antibunching, optical stability, and compatibility with nanophotonic and plasmonic architectures. In particular, hBN defect centers and quantum dots demonstrated room-temperature single-photon emission and have recently been coupled to photonic and plasmonic nanocavities \cite{cavityhbn, qdcavity}, while TMD-based emitters often requiring cryogenic operation \cite{strainwese}.} A key prerequisite for exploiting these platforms is the reliable identification of isolated single-photon emitters, as opposed to unresolved multi-emitter clusters that compromise quantum performance.

Confocal microscopy is the conventional tool used to detect and locate single emitters by scanning and recording fluorescence intensity \cite{confocal}. In this method, a microscope objective illuminates the sample and collects the emitted fluorescence, which is directed to a detector. The sample, mounted on a piezo scanner, is then shifted to successive positions, and the fluorescence intensity is measured at each point. In this way, a complete fluorescence intensity map of the sample is constructed, with the brightness at each pixel corresponding to the fluorescence recorded at that position \cite{brightness}. Typically, regions of reduced brightness are then analyzed further by calculating the second-order correlation function, $g^{(2)}(0)$, to verify the presence of single emitters \cite{g2a}. However, the microscope objective collects light from the entire region within its focal spot—typically on the order of $1\,\mu$m—without providing information about how emitters are distributed within that spot \cite{confocal}.

This limitation is particularly critical for applications requiring precise emitter placement, such as coupling single-photon emitters to optical cavities, waveguides, or plasmonic structures \cite{plasmonNV,wavguideNV}. For example, efficient cavity coupling requires that a single emitter be positioned at the antinode of the cavity field \cite{antinode}. Standard confocal imaging can determine the number of emitters within a diffraction-limited spot but cannot resolve whether a true single emitter is located at the desired position within that spot.

By reconstructing the emitter distribution with sub-focal-spot precision, our approach enables experimentalists to identify suitable sites directly, thereby reducing the effort required to locate and couple single emitters. This capability is equally valuable in fabrication workflows, where feedback on emitter locations can guide the positioning of photonic structures such as ring resonators \cite{location_NVcavity}. Moreover, fabrication steps including ion implantation, annealing, and etching critically depend on knowledge of the spatial distribution of emitters \cite{implantation}. Reliable localization of single-photon emitters thus accelerates the development of scalable solid-state quantum photonic devices across a broad range of material platforms.

In this work, we address these challenges by simulating a raster-scanned $g^{(2)}(0)$ mapping technique combined with a reconstruction algorithm.  {Using synthetic photon-timestamp data generated from a forward model of randomly distributed single-photon emitters under a scanned Gaussian excitation profile}, rather than relying on fluorescence intensity alone, we compute local photon antibunching across the field of view, which encodes the effective number of emitters within each focal spot. By iteratively inverting this forward model, we reconstruct the emitter occupancy on a sub-diffraction-limited grid.  { Although this work uses NV centers as representative systems for single photon emitters, the reconstruction method is not specific to NV centers as it relies solely on spatially resolved measurements of the second-order correlation function $g^{(2)}(0)$, which is a universal signature of single-photon emission and is independent of the microscopic nature of the emitter and is, in principle, applicable to a wide class of solid-state single-photon sources, including hBN defect centers, TMD emitters, and quantum dots. Different material systems will demand different experimental conditions (such as, for instance, cryogenic temperatures), which are not associated with the algorithm per se, which only works with a $g^{(2)}(0)$ scan. }

This approach not only distinguishes isolated single emitters from clustered multi-emitter sites, but also provides their spatial distribution within regions smaller than the diffraction-limited collection area, thereby enabling reliable identification and localization of true single-photon sources  {in simulation and in a manner that is directly transferable to a broad class of solid-state emitter systems}. Altogether, we believe these  {simulation results} establish our method as a powerful diagnostic and design tool for advancing solid-state quantum photonic technologies.

\section{The Algorithm}

The second-order intensity correlation function is defined as \cite{g2}
\begin{equation}
\label{eq:1}
g^{(2)}(\tau)=\frac{\langle I(t)\, I(t+\tau) \rangle}{\langle I(t)\rangle^{2}},
\end{equation}
where \(I(t)\) denotes the detected intensity at the detector.
In a photon-counting Hanbury Brown–Twiss (HBT) experiment, an equivalent estimator is
\begin{equation}
\label{eq:2}
g^{(2)}(\tau)=\frac{\langle n_{1}(t)\,n_{2}(t+\tau)\rangle}{\langle n_{1}(t)\rangle\,\langle n_{2}(t)\rangle},
\end{equation}
with \(n_{1,2}(t)\) the counts per bin \(\Delta t\) in the two channels. An ideal single
photon emitter exhibits antibunching with $g^{(2)}(0)=0$. For \(N\) identical, independent emitters detected with equal brightness, the zero-delay value is
$g^{(2)}(0)=1-\frac{1}{N}$,
which we invert to obtain $N_{\mathrm{meas}}=1/\!\big(1-g^{(2)}(0)\big)$.
More generally, if emitters contribute with non-negative weights \(w_j\) owing to non-uniform pump distribution,
\begin{equation}
\label{eq:g2_weighted}
g^{(2)}(0)=1-\frac{\sum_{i} w_i^{2} n_i}{\Big(\sum_{i} w_i n_i\Big)^{2}}
\quad\Longleftrightarrow\quad
N_{\mathrm{eff}}=\frac{\Big(\sum_{i} w_i n_i\Big)^{2}}{\sum_{i} w_i^{2} n_i}\,,
\end{equation}
where \(n_i\) is the (integer) number of emitters in pixel \(i\) and \(N_{\mathrm{eff}}\) is
the effective emitter number sensed by the focal spot.

\begin{figure}[t]
  \centering
    \begin{overpic}[width=0.49\textwidth]{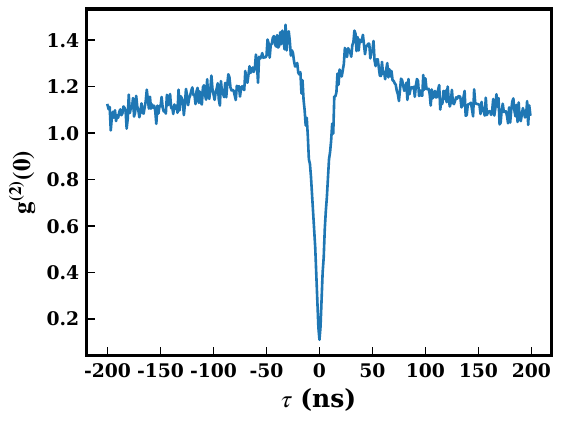}
    \put(-3,70){\small\bfseries (a)}
    \phantomsubcaption\label{fig1a}
  \end{overpic}\hfill
    \begin{overpic}[width=0.49\textwidth]{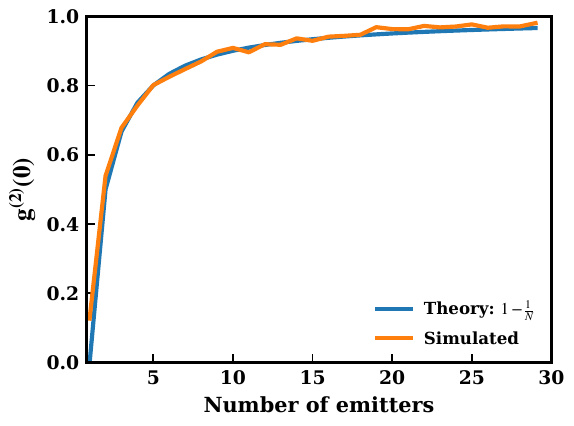}
    \put(-3,70){\small\bfseries (b)}
    \phantomsubcaption\label{fig1b}
  \end{overpic}
  \caption{Simulated results. 
  (a) $g^{(2)}(\tau)$ for a single NV center in diamond, showing antibunching with $g^{(2)}(0)=0.11$. 
  (b) $g^{(2)}(0)$ as a function of the number of emitters $N$: theory $g^{(2)}(0)=1-\tfrac{1}{N}$ (blue) and simulations (orange).}
\end{figure}

 {
To simulate emission from a single NV center including intersystem crossing
(ISC), we model the emitter as a three-level system with ground ($G$), excited
($E$), and metastable shelving ($S$) states. Optical pumping drives transitions
$G \rightarrow E$ at rate $k_{ge}$, while the excited state decays either
radiatively ($E \rightarrow G$) at rate $\gamma$ or non-radiatively via ISC
($E \rightarrow S$) at rate $k_{es}$. Population shelved in $S$ returns to the
ground state at rate $k_{sg}$. Photon emission events correspond to the
radiative $E \rightarrow G$ transitions. The resulting inter-photon delays
arise from this stochastic excitation--decay cycle, in which intermittent
population of the metastable shelving state introduces dark periods between
emission events, leading to characteristic photon bunching at longer time
delays.
} The resulting timestamps are split at random into two HBT channels (a virtual 50:50 beamsplitter),  {after which realistic effects are applied independently to each channel, including a detector dead time of 55~ns, background fluorescence (approximately $10\%$ of the detected counts in each channel), detector timing jitter of 400~ps, and a finite detection efficiency of 80\%.} 
The second-order correlation function $g^{(2)}(\tau)$ is then computed from Eq.~\ref{eq:2}, whose trace is shown in Fig \ref{fig1a}. 
A pronounced antibunching minimum at zero delay is observed with  { $g^{(2)}(0)=0.11$}, consistent with near-ideal single-photon emission. 
The small residual value at $\tau=0$ is attributed to  {finite counting statistics, the applied detector nonidealities, and the finite histogram bin width used to evaluate $g^{(2)}$.}  {In these simulations, the detected single-photon rate on each HBT
channel is of order $10^{6}$~cps. For a per-pixel acquisition time of $0.1\,\mathrm{s}$, this corresponds to total photon counts of order $10^{5}$ per detector channel and approximately $10^{3}$ coincidence events within a $1\,\mathrm{ns}$ wide zero-delay time bin, yielding the observed $g^{(2)}(0)$ value. This scenario is consistent with values reported for single NV centers in the literature~\cite{g2_exp}.}
 

Having established with Fig \ref{fig1a} that the simulator reproduces the single-emitter
antibunching signature, we next examine how \(g^{(2)}(0)\) scales with the number of
emitters \(N\).  {We generate $N$ independent photon timestamp streams by simulating the three-level excitation–decay
dynamics for each emitter.} The streams are concatenated and time-sorted to form a composite detection record, which
is then split at random into two HBT channels (virtual 50:50
beamsplitter)  {after adding all nonidealities} and $g^{(2)}(\tau)$ is computed from Eq. \ref{eq:2}. Fig \ref{fig1b} shows the resulting zero-delay correlation \(g^{(2)}(0)\) versus \(N\).
The simulations (orange) closely follow the theoretical prediction
\(g^{(2)}(0)=1-\tfrac{1}{N}\) for identical, independent emitters (blue).
As \(N\) increases, \(g^{(2)}(0)\) rises from \(\approx 0\) at \(N=1\) toward unity,
indicating the transition from single-photon antibunching to the Poissonian
multi-emitter limit. Minor discrepancies at small \(N\) were confirmed to arise from finite counting statistics.

Having established the quantitative link between \(g^{(2)}(0)\) and the emitter number \(N\) (Fig \ref{fig1b}), we now turn to reconstruction, i.e, inferring the underlying NV-center configuration from spatially resolved \(g^{(2)}(0)\) measurements, thereby recovering sub-spot structure.

\begin{figure}[ht]
  \centering
  \begin{subfigure}[b]{0.48\textwidth}
    \centering
    \includegraphics[width=\textwidth]{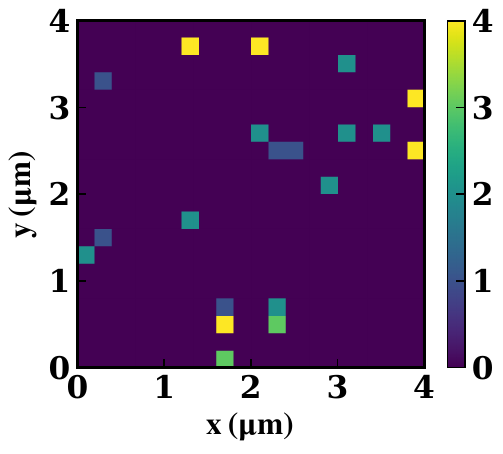}
    \caption{Ground truth.}
    \label{fig2a}
  \end{subfigure}
  \hfill
  \begin{subfigure}[b]{0.51\textwidth}
    \centering
    \includegraphics[width=\textwidth]{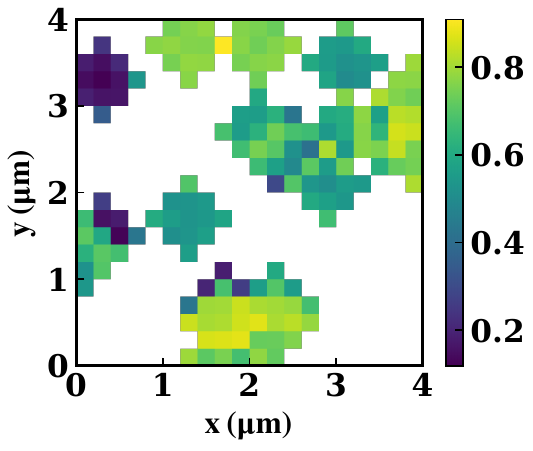}
    \caption{$g^{(2)}(0)$ map.}
    \label{fig2b}
  \end{subfigure}

  \caption{(a) shows Spatial distribution of NV centers over a \(4\times4~\mu\mathrm{m}^{2}\) sample area, represented on a 20\(\times\)20-pixel grid. The color scale indicates the number of emitters per pixel and (b) shows raster-scan map of \(g^{(2)}(0)\) over the NV-center distribution in a \(4\times4~\mu\mathrm{m}^{2}\) field. For each pixel, \(g^{(2)}(0)\) is computed from photons collected by an 800\,nm diameter focal spot centered on that pixel, as the focal spot is raster-scanned across the NV map. The color scale shows the resulting \(g^{(2)}(0)\). Pixels where the focal spot contains no NV centers are left undefined (NaN; white).}
  \label{fig:combined}
\end{figure}

 {In a $g^{(2)}(\tau)$ measurement}, a microscope objective collects fluorescence from all NV centers
excited within a diffraction-limited focal spot. The $\sim 800~\mathrm{nm}$ spot
 {is assumed in the simulation to excite} multiple emitters
simultaneously, so the recorded $g^{(2)}(0)$ reflects only the effective number
of emitters present in that region. To recover the underlying spatial distribution
within the spot,  {we numerically raster-scan the focal spot} to form
a simulated $g^{(2)}(0)$ map of the region. For validation, we simulate a reference distribution over a \(4\times4~\mu\mathrm{m}^{2}\) field discretized at 200\,nm (20\(\times\)20 pixels), containing 20 nonzero pixels with integer occupancies (1–4), as shown in Fig \ref{fig2a}. This reference map, i.e, the ground truth, is used to generate the \(g^{(2)}(0)\) map and to assess reconstruction accuracy.

Having linked \(g^{(2)}(0)\) to emitter number, we next build a spatially resolved \(g^{(2)}(0)\) map to serve as input for reconstruction. At each scan position (pixel), an $800\,\mathrm{nm}$ diameter focal spot is centered on that pixel. Pixels $j$ that lie under the focal spot are assigned Gaussian weights
\( w_j = \exp\left[-\frac{d_j^{2}}{2\sigma^{2}}\right],
\) with $\sigma = r/2 = 0.2~\mu\mathrm{m}$ ($r=0.4~\mu\mathrm{m}$). With a $200\,\mathrm{nm}$ pixel pitch, this corresponds to a radius of two pixels, so roughly $\pi r^{2}/\Delta^{2}\approx 12$ pixels ($\Delta=0.2~\mu\mathrm{m}$) contribute to the signal.  {Photon detection is simulated} by retaining each photon from
pixel $j$ independently with probability $p_j = w_j$ (e.g.,
$w_j=0.2 \Rightarrow 20\%$ of timestamps are kept), noting that the Gaussian ensures
$0 \leq w_j \leq 1$. The retained timestamps from all contributing pixels are
concatenated and time-sorted to form the
cpmposite photon record for that scan position. This record is
then split into two HBT channels by a virtual 50:50 beamsplitter
to compute $g^{(2)}(0)$. When the focal spot is shifted to the next scan position (a $200\,\mathrm{nm}$ step), the new spot again excites and collects from all pixels under its $800\,\mathrm{nm}$ footprint. As a result, some of the same pixels contribute as in the previous position (but with different weights, reflecting the shifted spot), while new pixels also enter the illuminated region.  By repeating this procedure over all scan positions,  {we numerically construct} the complete
raster-scanned $g^{(2)}(0)$ map.

Figure \ref{fig2b} presents the resulting \(g^{(2)}(0)\) map over a \(4\times4~\mu\mathrm{m}^{2}\) field. The color scale reports the locally evaluated \(g^{(2)}(0)\); white regions indicate scan positions where the focal spot contains no NV centers and the statistic is undefined (NaN).  {In particular, in regions between NV centers where the detected photon counts and coincidence rates fall below a predefined threshold (above background), $g^{(2)}(0)$ is not evaluated and
the corresponding pixels are set to NaN to avoid unreliable estimates.}

\begin{figure}[ht]
    \centering

    \begin{subfigure}{0.48\linewidth}
        \centering       \includegraphics[width=\linewidth]{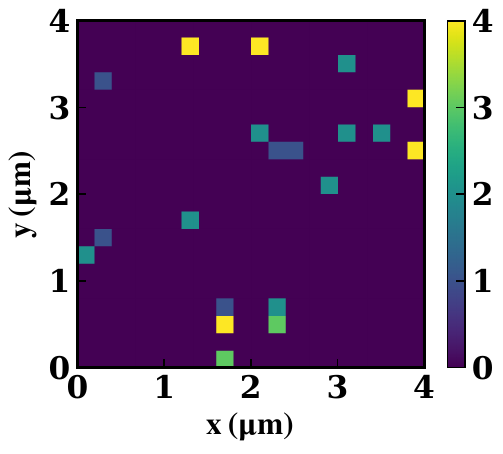}
        \caption{Reconstructed matrix.}
        \label{fig3a}
    \end{subfigure}\hfill
    \begin{subfigure}{0.5\linewidth}
        \centering       \includegraphics[width=\linewidth]{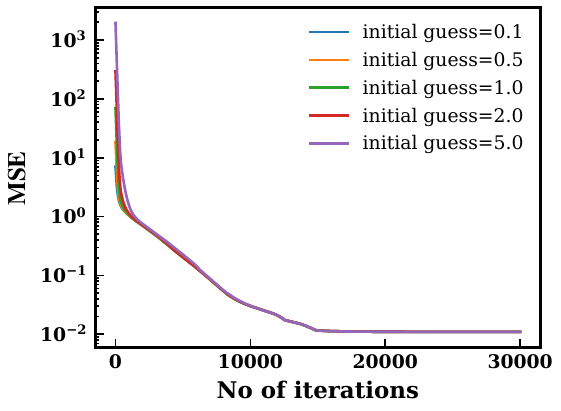}
        \caption{Convergence: MSE vs iterations for different initial guesses}
        \label{fig3b}
    \end{subfigure}

    \vspace{0.6em}
    \begin{subfigure}{0.48\linewidth}
        \centering      \includegraphics[width=\linewidth]{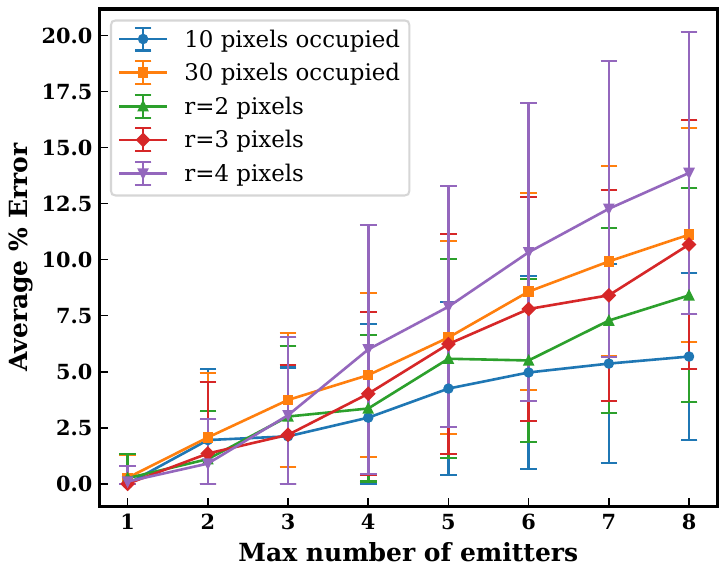}
        \caption{Error in reconstruction.}
        \label{fig3c}
    \end{subfigure}
    \caption{(a) shows reconstructed NV-center distribution obtained by using the \(g^{(2)}(0)\) map of Fig.~4. The color scale indicates the number of NV centers per pixel. (b) Mean squared error (MSE) versus iteration number for different initial guesses, showing robust convergence of the reconstruction algorithm and (c) shows error in reconstructing the original NV center distribution from its \(g^{(2)}(0)\) map, plotted versus the maximum number of NV centers per pixel. Error bars indicate the standard deviation across realizations.}
    \label{fig3}
\end{figure}

For each scan position $\mathbf r_0$ we convert the measured zero-delay
second-order correlation to an effective emitter number
\[
N_{\mathrm{meas}}(\mathbf r_0)
=
\frac{1}{1-g^{(2)}_{\mathrm{meas}}(0;\mathbf r_0)},
\qquad 0 \le g^{(2)}_{\mathrm{meas}} < 1 .
\]

With Gaussian weights $w_i(\mathbf r_0)$ defined over pixels $i$, we introduce
\[
A(\mathbf r_0)=\sum_i w_i(\mathbf r_0)\, n_i,\qquad
B(\mathbf r_0)=\sum_i w_i^2(\mathbf r_0)\, n_i,
\]
 {and the predicted effective emitter number}
\[
N_{\mathrm{eff}}(\mathbf r_0)=\frac{A(\mathbf r_0)^2}{B(\mathbf r_0)},
\]
where $n_i \ge 0$ denotes the unknown NV-center occupancy in pixel $i$.

 {We seek an occupancy map $\{n_i\}$ that is consistent with the measured
effective numbers across all valid scan positions by minimizing the global
least-squares objective}
\[
\mathcal{L}(\{n_i\})
=
\sum_{\mathbf r_0 \in \mathrm{scan}}
\Big[
N_{\mathrm{meas}}(\mathbf r_0)
-
N_{\mathrm{eff}}(\mathbf r_0;\{n_i\})
\Big]^2 ,
\]
 {where scan positions with undefined or invalid $g^{(2)}$ values are excluded.}

 {Beginning with an initial estimate $\{n_i^{(0)}\}$ the focal spot is raster-scanned across the image, and the occupancies of pixels within the Gaussian support are updated iteratively using the local residual at each scan position.}

\[
\varepsilon(\mathbf r_0)
=
N_{\mathrm{meas}}(\mathbf r_0)
-
N_{\mathrm{eff}}(\mathbf r_0),
\]
and update only those pixels $i$ for which $w_i(\mathbf r_0)\neq 0$.

The derivative of $N_{\mathrm{eff}}$ with respect to $n_i$ is
\[
\frac{\partial N_{\mathrm{eff}}(\mathbf r_0)}{\partial n_i}
=
\frac{
2A(\mathbf r_0)\,w_i(\mathbf r_0)\,B(\mathbf r_0)
-
A(\mathbf r_0)^2\,w_i(\mathbf r_0)^2
}{
B(\mathbf r_0)^2
}.
\]

The occupancies are updated according to
\[
n_i
\;\leftarrow\;
\max\!\Big\{
0,\;
n_i
+
\lambda\,\varepsilon(\mathbf r_0)\,
\frac{\partial N_{\mathrm{eff}}(\mathbf r_0)}{\partial n_i}
\Big\},
\]
 {where $\lambda$ is a fixed learning rate controlling the step size of the
updates. Each scan position updates only nearby pixels; however, because the focal spots overlap and the scan is repeated, information from all measurements is progressively combined, leading to a solution that is consistent with the global least-squares objective. The procedure is repeated for multiple full raster sweeps until convergence, which in practice is determined either by a fixed number of iterations or by saturation of the mean squared residual. The learning rate is chosen to be sufficiently small (\(\lambda = 10^{-4}\) in all simulations reported here) to ensure stable convergence of the optimization \cite{iterations}}.

After convergence, the continuous values $\{n_i\}$ are projected onto integers,
yielding the reconstructed NV-center occupancy map shown in
Fig \ref{fig3a}.  {Robustness to initialization was verified by repeating the
reconstruction for multiple uniform initial guesses
($n_i^{(0)} \in \{0.1, 0.5, 1, 2, 5\}$), as shown in Fig \ref{fig3b}. All these initial guesses yielded consistent converged occupancy map of Fig \ref{fig3a}.}

Having reconstructed the occupancy map (Fig \ref{fig3a}),  {from simulated data}, we next quantify the accuracy of the method.

Fig \ref{fig3c} reports the mean percentage error in reconstructing the original 
NV-center distribution from its $g^{(2)}(0)$ map as a function of the maximum 
per-pixel occupancy $N_{\max}$. Each marker denotes the mean over multiple random 
realizations, and error bars represent one standard deviation. The error remains 
small for sparse scenes ($N_{\max}\!\lesssim\!3$) and increases monotonically with 
$N_{\max}$. At low occupancies, the theoretical levels $g^{(2)}(0)=1-\tfrac{1}{N}$ 
are well separated, as seen in Fig \ref{fig1b}; the spacing between successive $N$ 
is $\Delta g^{(2)}(0)=\tfrac{1}{N(N+1)}$, so reconstruction errors remain small. 
As $N$ increases these levels compress toward unity, reducing separability in the 
presence of noise and binning, and thereby driving the larger errors observed at 
high $N$.

The green curve in Fig \ref{fig3c} shows the reconstruction error for the case 
where the number of nonzero elements is fixed at 20 and the focal-spot radius is 
$r=2$ pixels ($400~\mathrm{nm}$), with a raster scan step size of $200~\mathrm{nm}$, 
which sets the reconstruction resolution. The average percentage error is plotted 
as a function of $N_{\max}$. The red and purple curves report the reconstruction 
error when the same scene is raster-scanned with finer step sizes of 
$133~\mathrm{nm}$ ($r=3$ pixels) and $100~\mathrm{nm}$ ($r=4$ pixels), respectively, 
thereby probing emitter locations on a denser grid.  {The error trends across $200~\mathrm{nm}$, $133~\mathrm{nm}$, and $100~\mathrm{nm}$ sampling show a systematic increase in reconstruction error as the scan step is reduced below the focal-spot scale. At the same time, finer sampling improves the achievable spatial resolution, illustrating the inherent trade-off between resolution and reconstruction accuracy under increasingly dense scanning.}

The blue and orange curves in Fig \ref{fig3c} show the reconstruction error as 
the overall emitter density in the field is varied. Specifically, the blue curve 
corresponds to scenarios containing 10 occupied pixels (nonzero elements), while the 
orange curve corresponds to 30 occupied pixels. The similarity of the error 
trends confirms that the reconstruction algorithm remains robust as the 
emitter density in the sample is increased. In all cases, markers indicate the mean across realizations, and error bars denote one standard deviation.

\begin{figure}[ht]
\makebox[\textwidth][l]{\includegraphics[width=1\textwidth]{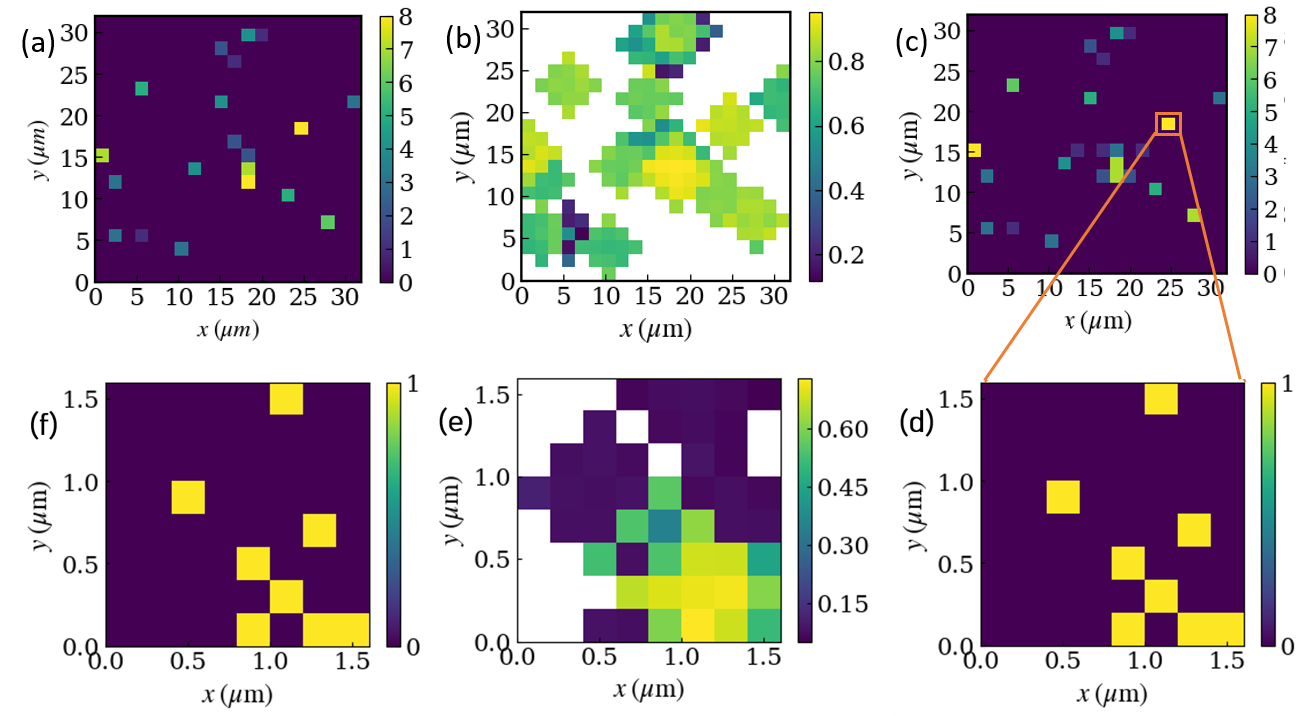}}
\caption{(a) Original NV-center distribution over a \(32\times32~\mu\mathrm{m}^{2}\) field sampled at \(1.6~\mu\mathrm{m}\) per pixel. 
(b) \(g^{(2)}(0)\) map obtained by raster-scanning (a) with a 6.4\,\(\mu\)m diameter focal spot (low magnification). 
(c) Emitter distribution reconstructed from (b). 
(d) Ground-truth emitter distribution within a selected \(1.6\times1.6~\mu\mathrm{m}^{2}\) pixel. 
(e) Local \(g^{(2)}(0)\) map of that tile acquired with an 800\,nm diameter focal spot (high magnification). 
(f) Reconstruction from (e).}
\label{fig4}
\end{figure}

We now utilise the reconstruction algorithm to identify the location of a single emitter or a few emitters in a subfocal region. A coarse $g^{(2)}(0)$ map acquired with an objective that has a 6.4\,\(\mu\)m focal spot is inverted to estimate the occupancy in each coarse pixel. We then confine the high-magnification scan (objective with 800\,nm focal spot) to these regions of interest (ROIs).  {Importantly, the coarse stage is used only for ROI selection and does not determine the final emitter position; therefore, errors in the coarse map do not propagate to the second stage, since the final emitter-number estimation and localisation are performed independently from the fine-scan data.} Alternatively,
This multi-resolution strategy enables accurate localisation with higher resolution, improved SNR, and  {reduced} acquisition time.
 {This strategy can be implemented as an augmented
intensity-scanning scheme, in which conventional intensity imaging is used to identify regions of low intensity or ambiguous contrast, while the reconstruction algorithm is selectively applied within these regions to
resolve emitter-number occupancy inside subfocal areas.}

Fig \ref{fig4} shows the recommended experimental procedure for locating a single NV center.
(a) shows ground-truth NV-center distribution over a \(32\times32~\mu\mathrm{m}^{2}\) field sampled at \(1.6~\mu\mathrm{m}\) per pixel, colors indicate the integer number of NV centers per pixel. (b) shows coarse \(g^{(2)}(0)\) map obtained by raster-scanning (a) with a 6.4\,\(\mu\)m diameter focal spot (lower numerical aperture (NA)). (c) shows emitter distribution reconstructed from (b) where the highlighted tile is selected as a region of interest for single-NV search. (d) shows ground-truth NV-center distribution within the selected \(1.6\times1.6~\mu\mathrm{m}^{2}\) area sampled at \(200~\mathrm{nm}\) per pixel. (e) shows local \(g^{(2)}(0)\) map of that ROI acquired by raster-scanning with an 800\,nm diameter
Gaussian focal spot (high NA) and (f) shows reconstruction of emitter distribution from (e).

\begin{figure}[t]
    \centering
    \begin{subfigure}[t]{0.32\textwidth}
        \centering
        \includegraphics[width=\linewidth]{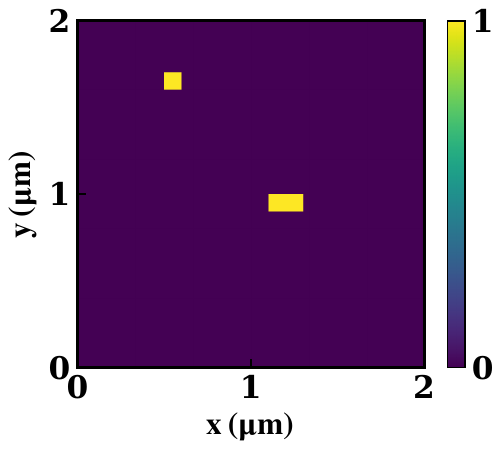}
        \caption{Ground truth.}
        \label{fig5a}
    \end{subfigure}
    \hfill
    \begin{subfigure}[t]{0.33\textwidth}
        \centering
        \includegraphics[width=\linewidth]{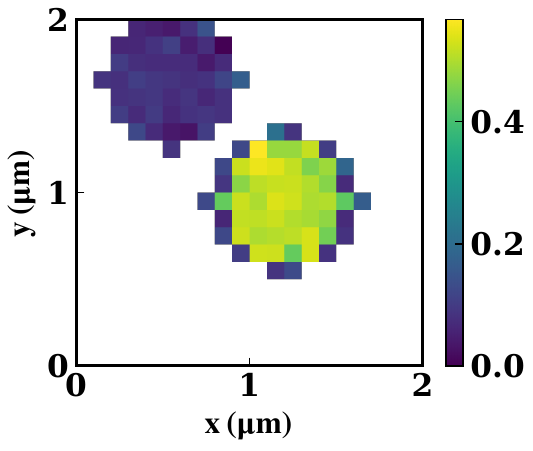}
        \caption{$g^{(2)}(0)$ map.}
        \label{fig5b}
    \end{subfigure}
    \hfill
    \begin{subfigure}[t]{0.33\textwidth}
        \centering
        \includegraphics[width=\linewidth]{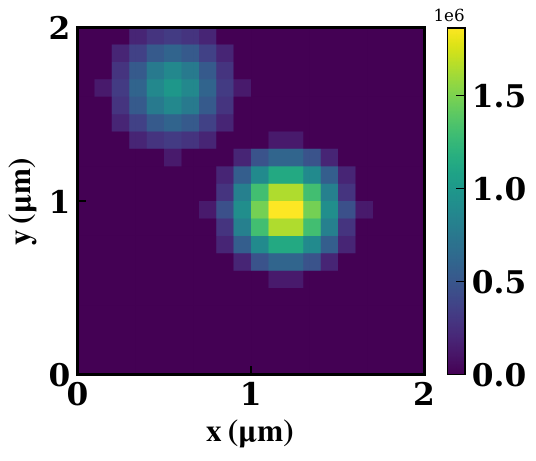}
        \caption{Intensity map.}
        \label{fig5c}
    \end{subfigure}
    
    \vspace{0.5em}
    
    \begin{subfigure}[t]{0.35\textwidth}
        \centering
        \includegraphics[width=\linewidth]{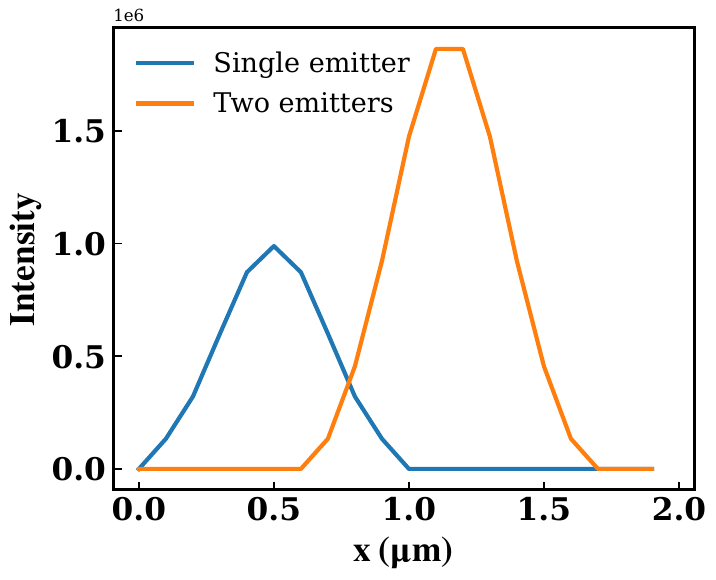}
        \caption{PSF.}
        \label{fig5d}
    \end{subfigure}
    \begin{subfigure}[t]{0.32\textwidth}
        \centering
        \includegraphics[width=\linewidth]{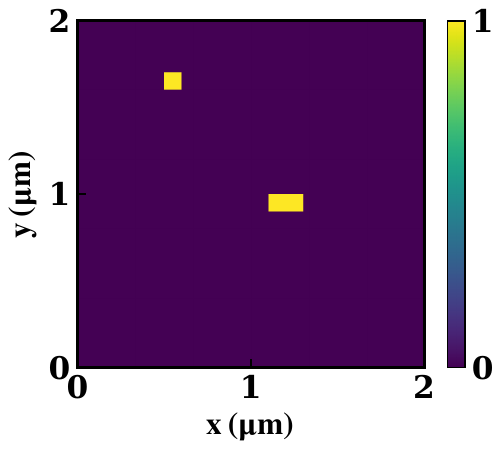}
        \caption{Reconstructed matrix.}
        \label{fig5e}
        
    \end{subfigure}

    \caption{Sub-diffraction localization and resolution using photon-correlation reconstruction. (a) Ground-truth configuration with a single emitter and two emitters separated by $100\,\mathrm{nm}$. (b) Corresponding spatial $g^{(2)}(0)$ map. (c) Intensity map from conventional raster scanning, exhibiting diffraction-limited fluorescence spots that do not resolve closely spaced emitters. (d) Extracted one-dimensional PSF profiles for a single emitter and for two emitters separated by 100~nm. (e) Reconstructed occupancy map, which accurately localizes the single emitter and resolves the two closely spaced emitters.}

    \label{fig5}
\end{figure}

 {
To demonstrate the super-resolution capability of our reconstruction algorithm, we consider a single simulation framework in which both single-emitter localization and two-emitter resolution are examined. Fig \ref{fig5a} shows the ground-truth configuration, including a single emitter and a pair of emitters separated by $100\,\mathrm{nm}$. The corresponding spatial $g^{(2)}(0)$ map is shown in Fig \ref{fig5b}. For comparison, Fig \ref{fig5c} shows the conventional intensity map obtained from raster scanning. As expected, the intensity distribution is governed by the diffraction-limited point-spread function and fails to resolve the two emitters separated by $100\,\mathrm{nm}$. This is further illustrated in Fig \ref{fig5d}, which shows one-dimensional PSF profiles for a single emitter and for two emitters, demonstrating that increased intensity does not translate into spatial resolvability below the diffraction limit.
}

 {Applying our reconstruction algorithm to the same $g^{(2)}(0)$ data yields the reconstructed occupancy map shown in Fig \ref{fig5e}. The algorithm accurately localizes the single emitter and clearly resolves the two emitters separated by $100\,\mathrm{nm}$, demonstrating localization precision and spatial resolution beyond the diffraction limit. These results are obtained by $800\,\mathrm{nm}$ focal spot and a raster scan step size of $100\,\mathrm{nm}$.
}

 {Our method encodes emitter-number information that is inaccessible to
conventional intensity-based imaging \cite{emitterno} We note that SOFI
(super-resolution optical fluctuation imaging) and higher-order
correlation microscopy~\cite{SOFI,Schwartz2013} achieve super-resolution by
effectively narrowing the point-spread function through intensity or
correlation cumulants, reaching lateral resolutions as fine as
$55 \pm 3\,\mathrm{nm}$ in high-order SOFI. In contrast,
antibunching-based and photon-statistics-based approaches~\cite{Israel2017},
including the present work, achieve sub-diffraction resolution by
discriminating and localizing emitters using nonclassical photon
statistics rather than by PSF narrowing; for example, antibunching has
been used to reliably separate and localize two quantum dots spaced by
approximately $100\,\mathrm{nm}$.}

\begin{figure}[ht]
\makebox[\textwidth][l]{\includegraphics[width=1\textwidth]{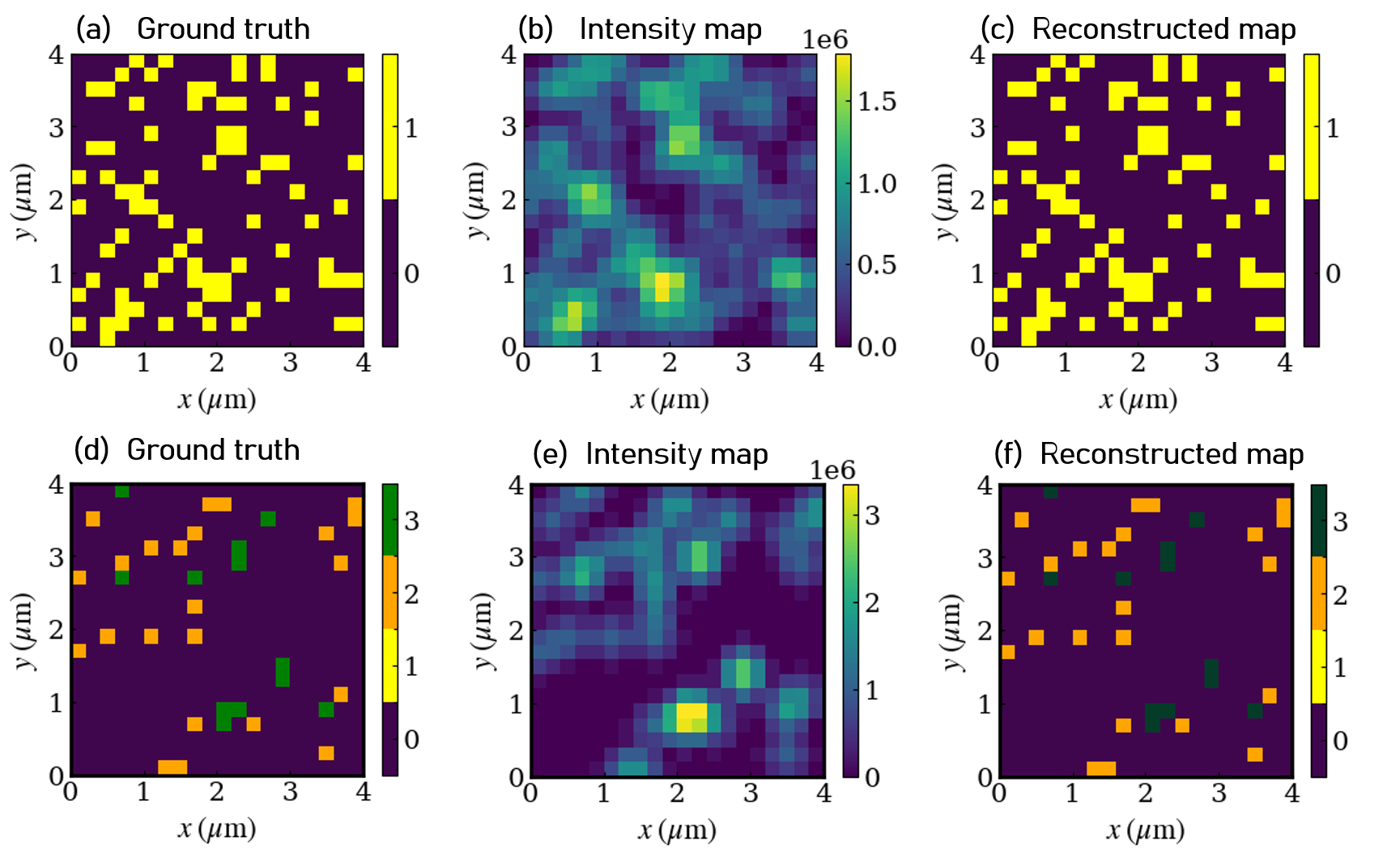}}
\caption{Comparison of conventional intensity mapping and reconstruction. 
  (a) Ground-truth NV-center distribution within a $4\times 4~\mu\text{m}^2$ region containing only single emitters. 
  (b) Intensity map obtained by raster-scanning the region with a focal spot. 
  (c) Reconstructed occupancy map from (b), which correctly reproduces single-emitter pixels. 
  (d) Ground-truth distribution for a region containing only multi-emitter sites, but no single emitters. 
  (e) Corresponding intensity map from raster scanning. The weakest intensity locations need to be studied further in the conventional technique. 
  (f) Reconstruction using our algorithm, which correctly identifies the absence of isolated single emitters.}
\label{fig6}
\end{figure}

Having established  {in simulation} that the reconstruction algorithm can resolve the number and location 
of emitters, we now present two scenarios that clearly demonstrate how reconstruction 
outperforms conventional intensity mapping. In particular, reconstruction reduces 
experimental effort by preventing the rejection of usable single-photon sites and by 
avoiding wasted searches in regions that do not contain single emitters.

In Fig \ref{fig6}(a--c) represents a comparison between intensity map and our reconstruction algorithm for a scenario where sample contains only single NV centers, each separated by at least $200~\mathrm{nm}$. A conventional intensity map of this region Fig \ref{fig5}(b) shows several bright clusters that would typically be interpreted as multi-emitter sites, since intensity alone cannot discriminate between a true cluster and a collection of nearby single emitters within the focal spot. Our reconstruction method Fig \ref{fig6}(c), based on evaluating local $g^{(2)}(0)$ and inverting the forward model, correctly identifies these as spatially separated single emitters. This capability is of direct experimental relevance: in device-fabrication workflows where one seeks to couple a single NV center to a cavity or waveguide, conventional intensity mapping would suggest discarding such regions as unsuitable. In contrast, Figs \ref{fig6}(d–f) illustrate a case where the sample contains no single emitters, but only sites with multiple NV centers. Here, the corresponding intensity map Fig \ref{fig6}(e) still displays moderately bright regions due to relative signal variations, which could mislead an experimentalist into searching for single emitters. Our reconstruction Fig \ref{fig6}(f), however, directly reveals the absence of single emitters, ensuring that time and resources are not wasted on unproductive regions. In this way, reconstruction provides a decisive advantage over conventional intensity mapping by faithfully reporting the underlying single-emitter distribution.

\section{Conclusion}

In conclusion, we have demonstrated  {through numerical simulations} that raster-scanned $g^{(2)}(0)$ mapping 
combined with inversion-based reconstruction can reliably recover both the number and spatial distribution of NV centers within regions smaller than the confocal focal spot. Simulations confirm robust performance across varying emitter densities and different scanning resolutions. This method  {is proposed as a simulation-validated strategy} to locate true single-photon sources and to guide nanophotonic device fabrication. For example, in ring resonators and Fabry--Perot microcavities, the optical field has successive nodes or successive antinodes separated by only $\sim$ 130\,nm, and only NV centers placed close to a cavity antinode couple strongly, while others behave like background emitters and reduce the observed Purcell factor \cite{ringres,cavityfield}. In this context, our algorithm  {demonstrates in simulation} a clear advantage: it can pinpoint the emitter position with $\sim$ 200\,nm  accuracy, compared to the $\sim$ 800\,nm  resolution limit (microscope objective of NA 0.8) of standard intensity maps. More broadly, such reconstruction can be combined with cavity mode simulations to guide fabrication and alignment in a range of photonic devices. This enables optimization of parameters such as ring radius, waveguide gap, or mirror spacing, ensuring maximal Purcell enhancement and reliable integration of single photon emitters into nanophotonic structures.

\section*{Acknowledgments}
We thank Rounak Chatterjee and Kiran Bajar for their critical reading of the manuscript, and Rounak Chatterjee for helpful discussions. We also express our gratitude to the Department of Atomic Energy, Government of India, for funding under Project Identification No.\ RTI4002, DAE OM No.\ 1303/1/2020/R\&D-II/DAE/5567. The authors declare no conflict of interest.

\bibliographystyle{ieeetr}
\bibliography{ref}      

@article{
doi:10.1126/science.1113394,
author = {B. Darquié  and M. P. A. Jones  and J. Dingjan  and J. Beugnon  and S. Bergamini  and Y. Sortais  and G. Messin  and A. Browaeys  and P. Grangier },
title = {Controlled Single-Photon Emission from a Single Trapped Two-Level Atom},
journal = {Science},
volume = {309},
number = {5733},
pages = {454-456},
year = {2005},
doi = {10.1126/science.1113394},
URL = {https://www.science.org/doi/abs/10.1126/science.1113394},
eprint = {https://www.science.org/doi/pdf/10.1126/science.1113394}}

@article{PhysRevLett.58.203,
  title = {Nonclassical radiation of a single stored ion},
  author = {Diedrich, Frank and Walther, Herbert},
  journal = {Phys. Rev. Lett.},
  volume = {58},
  issue = {3},
  pages = {203--206},
  numpages = {0},
  year = {1987},
  month = {Jan},
  publisher = {American Physical Society},
  doi = {10.1103/PhysRevLett.58.203},
  url = {https://link.aps.org/doi/10.1103/PhysRevLett.58.203}
}

@article{molecule,
  title = {Photon antibunching in the fluorescence of a single dye molecule trapped in a solid},
  author = {Basch\'e, Th. and Moerner, W. E. and Orrit, M. and Talon, H.},
  journal = {Phys. Rev. Lett.},
  volume = {69},
  issue = {10},
  pages = {1516--1519},
  numpages = {0},
  year = {1992},
  month = {Sep},
  publisher = {American Physical Society},
  doi = {10.1103/PhysRevLett.69.1516},
  url = {https://link.aps.org/doi/10.1103/PhysRevLett.69.1516}
}

@article{
quantumdot,
author = {P. Michler  and A. Kiraz  and C. Becher  and W. V. Schoenfeld  and P. M. Petroff  and Lidong Zhang  and E. Hu  and A. Imamoglu },
title = {A Quantum Dot Single-Photon Turnstile Device},
journal = {Science},
volume = {290},
number = {5500},
pages = {2282-2285},
year = {2000},
doi = {10.1126/science.290.5500.2282},
URL = {https://www.science.org/doi/abs/10.1126/science.290.5500.2282},
eprint = {https://www.science.org/doi/pdf/10.1126/science.290.5500.2282}}

@article{qcomputation,
doi = {10.1088/0953-8984/18/21/S08},
url = {https://dx.doi.org/10.1088/0953-8984/18/21/S08},
year = {2006},
month = {may},
publisher = {},
volume = {18},
number = {21},
pages = {S807},
author = {Wrachtrup, Jörg and Jelezko, Fedor},
title = {Processing quantum information in diamond},
journal = {Journal of Physics: Condensed Matter}
}

@article{qcommunication,
  title = {Fault-Tolerant Quantum Communication Based on Solid-State Photon Emitters},
  author = {Childress, L. and Taylor, J. M. and S\o{}rensen, A. S. and Lukin, M. D.},
  journal = {Phys. Rev. Lett.},
  volume = {96},
  issue = {7},
  pages = {070504},
  numpages = {4},
  year = {2006},
  month = {Feb},
  publisher = {American Physical Society},
  doi = {10.1103/PhysRevLett.96.070504},
  url = {https://link.aps.org/doi/10.1103/PhysRevLett.96.070504}
}

@article{meterology,
  title={Nanoscale imaging magnetometry with diamond spins under ambient conditions},
  author={Balasubramanian, Gopalakrishnan and Chan, Ian Y. and Kolesov, Roman and Al-Hmoud, Mahmoud and Tisler, Joerg and Shin, Chang and Kim, Chang and Wojcik, Adam and Hemmer, Philip R. and Krueger, Anke and Hanke, Thomas and Leitenstorfer, Alfred and Bratschitsch, Rudolf and Jelezko, Fedor and Wrachtrup, J{\"o}rg},
  journal={Nature},
  volume={455},
  number={7213},
  pages={648--651},
  year={2008},
  publisher={Nature Publishing Group},
  doi={10.1038/nature07278},
  pmid={18833276}
}

@article{coherencetime,
  author    = {Balasubramanian, Gopalakrishnan and Neumann, Philipp and Twitchen, Daniel and Markham, Matthew and Kolesov, Roman and Mizuochi, Norikazu and Isoya, Junichi and Achard, Jocelyn and Beck, Johannes and Tissler, Julia and Jacques, Vincent and Hemmer, Philip R. and Jelezko, Fedor and Wrachtrup, J{\"o}rg},
  title     = {Ultralong spin coherence time in isotopically engineered diamond},
  journal   = {Nature Materials},
  volume    = {8},
  number    = {5},
  pages     = {383--387},
  year      = {2009},
  doi       = {10.1038/nmat2420},
  url       = {https://doi.org/10.1038/nmat2420},
  issn      = {1476-4660},
  publisher = {Nature Publishing Group}
}

@article{singlreadout,
  author    = {Irber, Dominik M. and Poggiali, Francesco and Kong, Fei and Kieschnick, Michael and L{\"u}hmann, Tobias and Kwiatkowski, Damian and Meijer, Jan and Du, Jiangfeng and Shi, Fazhan and Reinhard, Friedemann},
  title     = {Robust all-optical single-shot readout of nitrogen-vacancy centers in diamond},
  journal   = {Nature Communications},
  volume    = {12},
  number    = {1},
  pages     = {532},
  year      = {2021},
  doi       = {10.1038/s41467-020-20755-3},
  url       = {https://doi.org/10.1038/s41467-020-20755-3},
  issn      = {2041-1723}
}

@article{qimaging,
  title = {Sub-Rayleigh quantum imaging using single-photon sources},
  author = {Thiel, C. and Bastin, T. and von Zanthier, J. and Agarwal, G. S.},
  journal = {Phys. Rev. A},
  volume = {80},
  issue = {1},
  pages = {013820},
  numpages = {5},
  year = {2009},
  month = {Jul},
  publisher = {American Physical Society},
  doi = {10.1103/PhysRevA.80.013820},
  url = {https://link.aps.org/doi/10.1103/PhysRevA.80.013820}
}

@article{qip,
  author    = {Michael F{\"o}rtsch and Josef U. F{\"u}rst and Christoffer Wittmann and Dmitry Strekalov and Andrea Aiello and Maria V. Chekhova and Christine Silberhorn and Gerd Leuchs and Christoph Marquardt},
  title     = {A versatile source of single photons for quantum information processing},
  journal   = {Nature Communications},
  year      = {2013},
  volume    = {4},
  number    = {1},
  pages     = {1818},
  doi       = {10.1038/ncomms2838},
  url       = {https://doi.org/10.1038/ncomms2838},
  issn      = {2041-1723}
}

@article{HBN,
  author    = {Toan Trong Tran and Kerem Bray and Michael J. Ford and Milos Toth and Igor Aharonovich},
  title     = {Quantum emission from hexagonal boron nitride monolayers},
  journal   = {Nature Nanotechnology},
  year      = {2016},
  volume    = {11},
  number    = {1},
  pages     = {37--41},
  doi       = {10.1038/nnano.2015.242},
  url       = {https://doi.org/10.1038/nnano.2015.242},
  issn      = {1748-3395}
}

@article{SiV,
  author    = {Assegid Mengistu Flatae and Florian Sledz and Haritha Kambalathmana and Stefano Lagomarsino and Hongcai Wang and Nicla Gelli and Silvio Sciortino and Eckhard W{\"o}rner and Christoph Wild and Benjamin Butz and Mario Agio},
  title     = {Single-photon emission from silicon-vacancy color centers in polycrystalline diamond membranes},
  journal   = {Applied Physics Letters},
  year      = {2024},
  volume    = {124},
  number    = {9},
  pages     = {094001},
  doi       = {10.1063/5.0191665},
  url       = {https://doi.org/10.1063/5.0191665},
  issn      = {0003-6951}
}

@article{NV,
  title = {Photon-emission statistics for single nitrogen-vacancy centers},
  author = {Panadero, I. and Espin\'os, H. and Tsunaki, L. and Volkova, K. and Tobalina, A. and Casanova, J. and Acedo, P. and Naydenov, B. and Puebla, R. and Torrontegui, E.},
  journal = {Phys. Rev. Appl.},
  volume = {22},
  issue = {1},
  pages = {014035},
  numpages = {13},
  year = {2024},
  month = {Jul},
  publisher = {American Physical Society},
  doi = {10.1103/PhysRevApplied.22.014035},
  url = {https://link.aps.org/doi/10.1103/PhysRevApplied.22.014035}
}

@article{roomtemp,
  author    = {A. Beveratos and S. K{\"u}hn and R. Brouri and T. Gacoin and J.-P. Poizat and P. Grangier},
  title     = {Room temperature stable single-photon source},
  journal   = {The European Physical Journal D - Atomic, Molecular, Optical and Plasma Physics},
  year      = {2002},
  volume    = {18},
  number    = {2},
  pages     = {191--196},
  doi       = {10.1140/epjd/e20020023},
  url       = {https://doi.org/10.1140/epjd/e20020023},
  issn      = {1434-6079}
}

@article{initionalisation,
author = {Christopher G. Yale  and Bob B. Buckley  and David J. Christle  and Guido Burkard  and F. Joseph Heremans  and Lee C. Bassett  and David D. Awschalom },
title = {All-optical control of a solid-state spin using coherent dark states},
journal = {Proceedings of the National Academy of Sciences},
volume = {110},
number = {19},
pages = {7595-7600},
year = {2013},
doi = {10.1073/pnas.1305920110},
URL = {https://www.pnas.org/doi/abs/10.1073/pnas.1305920110},
eprint = {https://www.pnas.org/doi/pdf/10.1073/pnas.1305920110}}

@article{nvmagnetic,
doi = {10.1088/2058-9565/abffbd},
url = {https://dx.doi.org/10.1088/2058-9565/abffbd},
year = {2021},
month = {jun},
publisher = {IOP Publishing},
volume = {6},
number = {3},
pages = {034006},
author = {Lenz, Till and Wickenbrock, Arne and Jelezko, Fedor and Balasubramanian, Gopalakrishnan and Budker, Dmitry},
title = {Magnetic sensing at zero field with a single nitrogen-vacancy center},
journal = {Quantum Science and Technology}
}

@article{Plasmonics,
  title = {Controlled Coupling of a Single Nitrogen-Vacancy Center to a Silver Nanowire},
  author = {Huck, Alexander and Kumar, Shailesh and Shakoor, Abdul and Andersen, Ulrik L.},
  journal = {Phys. Rev. Lett.},
  volume = {106},
  issue = {9},
  pages = {096801},
  numpages = {4},
  year = {2011},
  month = {Feb},
  publisher = {American Physical Society},
  doi = {10.1103/PhysRevLett.106.096801},
  url = {https://link.aps.org/doi/10.1103/PhysRevLett.106.096801}
}

@article{Photonic,
  title = {Coupling of Nitrogen-Vacancy Centers to Photonic Crystal Cavities in Monocrystalline Diamond},
  author = {Faraon, Andrei and Santori, Charles and Huang, Zhihong and Acosta, Victor M. and Beausoleil, Raymond G.},
  journal = {Phys. Rev. Lett.},
  volume = {109},
  issue = {3},
  pages = {033604},
  numpages = {5},
  year = {2012},
  month = {Jul},
  publisher = {American Physical Society},
  doi = {10.1103/PhysRevLett.109.033604},
  url = {https://link.aps.org/doi/10.1103/PhysRevLett.109.033604}
}

@article{PC2,
  author    = {Dirk Englund and Brendan Shields and Kelley Rivoire and Fariba Hatami and Jelena Vučković and Hongkun Park and Mikhail D. Lukin},
  title     = {Deterministic Coupling of a Single Nitrogen Vacancy Center to a Photonic Crystal Cavity},
  journal   = {Nano Letters},
  year      = {2010},
  volume    = {10},
  number    = {10},
  pages     = {3922--3926},
  publisher = {American Chemical Society},
  doi       = {10.1021/nl101662v},
  url       = {https://doi.org/10.1021/nl101662v},
  issn      = {1530-6984}
}

@article{waveguide,
    author = {Fu, K.-M. C. and Santori, C. and Barclay, P. E. and Aharonovich, I. and Prawer, S. and Meyer, N. and Holm, A. M. and Beausoleil, R. G.},
    title = {Coupling of nitrogen-vacancy centers in diamond to a GaP waveguide},
    journal = {Applied Physics Letters},
    volume = {93},
    number = {23},
    pages = {234107},
    year = {2008},
    month = {12},
    issn = {0003-6951},
    doi = {10.1063/1.3045950},
    url = {https://doi.org/10.1063/1.3045950},
    eprint = {https://pubs.aip.org/aip/apl/article-pdf/doi/10.1063/1.3045950/13248200/234107_1_online.pdf},
}

@article{antinode,
  title = {Hybrid Nanocavity Resonant Enhancement of Color Center Emission in Diamond},
  author = {Barclay, Paul E. and Fu, Kai-Mei C. and Santori, Charles and Faraon, Andrei and Beausoleil, Raymond G.},
  journal = {Phys. Rev. X},
  volume = {1},
  issue = {1},
  pages = {011007},
  numpages = {7},
  year = {2011},
  month = {Sep},
  publisher = {American Physical Society},
  doi = {10.1103/PhysRevX.1.011007},
  url = {https://link.aps.org/doi/10.1103/PhysRevX.1.011007}
}

@article{g2,
  title = {Coherence measures for heralded single-photon sources},
  author = {Bocquillon, E. and Couteau, C. and Razavi, M. and Laflamme, R. and Weihs, G.},
  journal = {Phys. Rev. A},
  volume = {79},
  issue = {3},
  pages = {035801},
  numpages = {4},
  year = {2009},
  month = {Mar},
  publisher = {American Physical Society},
  doi = {10.1103/PhysRevA.79.035801},
  url = {https://link.aps.org/doi/10.1103/PhysRevA.79.035801}
}

@article{location_NVcavity,
  author    = {Dirk Englund and Brendan Shields and Kelley Rivoire and Fariba Hatami and Jelena Vučković and Hongkun Park and Mikhail D. Lukin},
  title     = {Deterministic coupling of a single nitrogen vacancy center to a photonic crystal cavity},
  journal   = {Nano Letters},
  year      = {2010},
  volume    = {10},
  number    = {10},
  pages     = {3922--3926},
  doi       = {10.1021/nl101662v},
  issn      = {1530-6984},
  url       = {https://doi.org/10.1021/nl101662v}
}

@article{implantation,
    author = {Sangtawesin, S. and Brundage, T. O. and Atkins, Z. J. and Petta, J. R.},
    title = {Highly tunable formation of nitrogen-vacancy centers via ion implantation},
    journal = {Applied Physics Letters},
    volume = {105},
    number = {6},
    pages = {063107},
    year = {2014},
    month = {08},
    issn = {0003-6951},
    doi = {10.1063/1.4892971},
    url = {https://doi.org/10.1063/1.4892971},
    eprint = {https://pubs.aip.org/aip/apl/article-pdf/doi/10.1063/1.4892971/13544740/063107_1_online.pdf},
}

@article{
confocal,
author = {A. Gruber  and A. Dräbenstedt  and C. Tietz  and L. Fleury  and J. Wrachtrup  and C. von Borczyskowski },
title = {Scanning Confocal Optical Microscopy and Magnetic Resonance on Single Defect Centers},
journal = {Science},
volume = {276},
number = {5321},
pages = {2012-2014},
year = {1997},
doi = {10.1126/science.276.5321.2012},
URL = {https://www.science.org/doi/abs/10.1126/science.276.5321.2012},
eprint = {https://www.science.org/doi/pdf/10.1126/science.276.5321.2012}}

@article{brightness,
author = {Rabeau, J. R. and Stacey, A. and Rabeau, A. and Prawer, S. and Jelezko, F. and Mirza, I. and Wrachtrup, J.},
title = {Single Nitrogen Vacancy Centers in Chemical Vapor Deposited Diamond Nanocrystals},
journal = {Nano Letters},
volume = {7},
number = {11},
pages = {3433-3437},
year = {2007},
doi = {10.1021/nl0719271},
    note ={PMID: 17902725},

URL = { 
      https://doi.org/10.1021/nl0719271
},
eprint = { 
    
        https://doi.org/10.1021/nl0719271
}
}

@article{g2a,
  title = {Stable Solid-State Source of Single Photons},
  author = {Kurtsiefer, Christian and Mayer, Sonja and Zarda, Patrick and Weinfurter, Harald},
  journal = {Phys. Rev. Lett.},
  volume = {85},
  issue = {2},
  pages = {290--293},
  numpages = {0},
  year = {2000},
  month = {Jul},
  publisher = {American Physical Society},
  doi = {10.1103/PhysRevLett.85.290},
  url = {https://link.aps.org/doi/10.1103/PhysRevLett.85.290}
}

@article{1sec,
  author    = {N. Bar-Gill and L. M. Pham and A. Jarmola and D. Budker and R. L. Walsworth},
  title     = {Solid-state electronic spin coherence time approaching one second},
  journal   = {Nature Communications},
  year      = {2013},
  volume    = {4},
  number    = {1},
  pages     = {1743},
  doi       = {10.1038/ncomms2771},
  url       = {https://doi.org/10.1038/ncomms2771},
  issn      = {2041-1723}
}

@article{ringres,
  author    = {Andrei Faraon and Paul E. Barclay and Charles Santori and Kai-Mei C. Fu and Raymond G. Beausoleil},
  title     = {Resonant enhancement of the zero-phonon emission from a colour centre in a diamond cavity},
  journal   = {Nature Photonics},
  year      = {2011},
  volume    = {5},
  number    = {5},
  pages     = {301--305},
  doi       = {10.1038/nphoton.2011.52},
  url       = {https://doi.org/10.1038/nphoton.2011.52},
  issn      = {1749-4893}
}

@article{cavityfield,
  title = {Purcell-Enhanced Single-Photon Emission from Nitrogen-Vacancy Centers Coupled to a Tunable Microcavity},
  author = {Kaupp, Hanno and H\"ummer, Thomas and Mader, Matthias and Schlederer, Benedikt and Benedikter, Julia and Haeusser, Philip and Chang, Huan-Cheng and Fedder, Helmut and H\"ansch, Theodor W. and Hunger, David},
  journal = {Phys. Rev. Appl.},
  volume = {6},
  issue = {5},
  pages = {054010},
  numpages = {10},
  year = {2016},
  month = {Nov},
  publisher = {American Physical Society},
  doi = {10.1103/PhysRevApplied.6.054010},
  url = {https://link.aps.org/doi/10.1103/PhysRevApplied.6.054010}
}

@article{plasmonNV,
  author    = {Esteban Bermúdez-Ureña and Carlos Gonzalez-Ballestero and Michael Geiselmann and Renaud Marty and Ilya P. Radko and Tobias Holmgaard and Yury Alaverdyan and Esteban Moreno and Francisco J. García-Vidal and Sergey I. Bozhevolnyi and Romain Quidant},
  title     = {Coupling of individual quantum emitters to channel plasmons},
  journal   = {Nature Communications},
  year      = {2015},
  volume    = {6},
  number    = {1},
  pages     = {7883},
  doi       = {10.1038/ncomms8883},
  url       = {https://doi.org/10.1038/ncomms8883},
  issn      = {2041-1723}
}

@article{wavguideNV,
author = {J. P. Hadden and V. Bharadwaj and B. Sotillo and S. Rampini and R. Osellame and J. D. Witmer and H. Jayakumar and T. T. Fernandez and A. Chiappini and C. Armellini and M. Ferrari and R. Ramponi and P. E. Barclay and S. M. Eaton},
journal = {Opt. Lett.},
number = {15},
pages = {3586--3589},
publisher = {Optica Publishing Group},
title = {Integrated waveguides and deterministically positioned nitrogen vacancy centers in diamond created by femtosecond laser writing},
volume = {43},
month = {Aug},
year = {2018},
url = {https://opg.optica.org/ol/abstract.cfm?URI=ol-43-15-3586},
doi = {10.1364/OL.43.003586},
}

@article{magneticimaging,
  title = {Magnetic-Field-Assisted Spectral Decomposition and Imaging of Charge States of $\mathrm{N}$-$V$ Centers in Diamond},
  author = {Chakraborty, T. and Bhattacharya, R. and Anjusha, V.S. and Nesladek, M. and Suter, D. and Mahesh, T.S.},
  journal = {Phys. Rev. Appl.},
  volume = {17},
  issue = {2},
  pages = {024046},
  numpages = {10},
  year = {2022},
  month = {Feb},
  publisher = {American Physical Society},
  doi = {10.1103/PhysRevApplied.17.024046},
  url = {https://link.aps.org/doi/10.1103/PhysRevApplied.17.024046}
}

@article{
qduppu,
author = {Ravitej Uppu  and Freja T. Pedersen  and Ying Wang  and Cecilie T. Olesen  and Camille Papon  and Xiaoyan Zhou  and Leonardo Midolo  and Sven Scholz  and Andreas D. Wieck  and Arne Ludwig  and Peter Lodahl },
title = {Scalable integrated single-photon source},
journal = {Science Advances},
volume = {6},
number = {50},
pages = {eabc8268},
year = {2020},
doi = {10.1126/sciadv.abc8268},
URL = {https://www.science.org/doi/abs/10.1126/sciadv.abc8268},
eprint = {https://www.science.org/doi/pdf/10.1126/sciadv.abc8268}}

@article{g2_exp,
  author    = {Chen, Xing and Greiner, Johannes N. and Wrachtrup, J{\"o}rg and Gerhardt, Ilja},
  title     = {Single Photon Randomness based on a Defect Center in Diamond},
  journal   = {Scientific Reports},
  year      = {2019},
  volume    = {9},
  number    = {1},
  pages     = {18474},
  doi       = {10.1038/s41598-019-54594-0},
  url       = {https://doi.org/10.1038/s41598-019-54594-0},
  issn      = {2045-2322}
}

@article{WSe2,
  author  = {Koperski, M. and Nogajewski, K. and Arora, A. and Cherkez, V. and Mallet, P. and Veuillen, J.-Y. and Marcus, J. and Kossacki, P. and Potemski, M.},
  title   = {Single photon emitters in exfoliated {WSe$_2$} structures},
  journal = {Nature Nanotechnology},
  volume  = {10},
  number  = {6},
  pages   = {503--506},
  year    = {2015},
  month   = jun,
  doi     = {10.1038/nnano.2015.67}
}

@article{cavityhbn,
title = "Plasmonic Nanocavity to Boost Single Photon Emission From Defects in Thin Hexagonal Boron Nitride",
author = "Mohammadjavad Dowran and Ufuk Kilic and Suvechhya Lamichhane and Adam Erickson and Joshua Barker and Mathias Schubert and Liou, \{Sy Hwang\} and Christos Argyropoulos and Abdelghani Laraoui",
note = "Publisher Copyright: {\textcopyright} 2024 The Author(s). Laser \& Photonics Reviews published by Wiley-VCH GmbH.",
year = "2025",
month = feb,
day = "5",
doi = "10.1002/lpor.202400705",
language = "English (US)",
volume = "19",
journal = "Laser and Photonics Reviews",
issn = "1863-8880",
publisher = "Wiley-VCH Verlag",
number = "3",
}

@article{qdcavity,
author = {Hoang, Thang B. and Akselrod, Gleb M. and Mikkelsen, Maiken H.},
title = {Ultrafast Room-Temperature Single Photon Emission from Quantum Dots Coupled to Plasmonic Nanocavities},
journal = {Nano Letters},
volume = {16},
number = {1},
pages = {270-275},
year = {2016},
doi = {10.1021/acs.nanolett.5b03724},
    note ={PMID: 26606001},

URL = { 
    
        https://doi.org/10.1021/acs.nanolett.5b03724
    
    

},
eprint = { 
    
        https://doi.org/10.1021/acs.nanolett.5b03724
    
    

}}

@article{Strainwese,
  author  = {Yu, Y. and Ge, J. and Luo, M. and Seo, I. C. and Kim, Y. and Eng, J. J. H. and Lu, K. and Wei, T. R. and Lee, S. W. and Gao, W. and Li, H. and Nam, D.},
  title   = {Dynamic Tuning of Single-Photon Emission in Monolayer {WSe$_2$} via Localized Strain Engineering},
  journal = {Nano Letters},
  volume  = {25},
  number  = {9},
  pages   = {3438--3444},
  year    = {2025},
  month   = mar,
  doi     = {10.1021/acs.nanolett.4c05450},
}

@article{Israel2017,
  author  = {Israel, Yonatan and Tenne, Ron and Oron, Dan and Silberberg, Yaron},
  title   = {Quantum correlation enhanced super-resolution localization microscopy enabled by a fibre bundle camera},
  journal = {Nature Communications},
  year    = {2017},
  volume  = {8},
  number  = {1},
  pages   = {14786},
  doi     = {10.1038/ncomms14786},
  issn    = {2041-1723},
}

@article{SOFI,
author = {T. Dertinger  and R. Colyer  and G. Iyer  and S. Weiss  and J. Enderlein },
title = {Fast, background-free, 3D super-resolution optical fluctuation imaging (SOFI)},
journal = {Proceedings of the National Academy of Sciences},
volume = {106},
number = {52},
pages = {22287-22292},
year = {2009},
doi = {10.1073/pnas.0907866106},
URL = {https://www.pnas.org/doi/abs/10.1073/pnas.0907866106},
eprint = {https://www.pnas.org/doi/pdf/10.1073/pnas.0907866106}}

@article{iterations,
author = {Candes, Emmanuel J. and Xiaodong Li and Soltanolkotabi, Mahdi},
title = {Phase Retrieval via Wirtinger Flow: Theory and Algorithms},
year = {2015},
issue_date = {April 2015},
publisher = {IEEE Press},
volume = {61},
number = {4},
issn = {0018-9448},
url = {https://doi.org/10.1109/TIT.2015.2399924},
doi = {10.1109/TIT.2015.2399924},
journal = {IEEE Trans. Inf. Theor.},
month = apr,
pages = {1985–2007},
numpages = {23}
}

@article{emitterno,
  title = {Time-Resolved Line Shapes of Single Quantum Emitters via Machine Learned Photon Correlations},
  author = {Proppe, Andrew H. and Lee, Kin Long Kelvin and Kaplan, Alexander E. K. and Ginterseder, Matthias and Krajewska, Chantalle J. and Bawendi, Moungi G.},
  journal = {Phys. Rev. Lett.},
  volume = {131},
  issue = {5},
  pages = {053603},
  numpages = {6},
  year = {2023},
  month = {Aug},
  publisher = {American Physical Society},
  doi = {10.1103/PhysRevLett.131.053603},
  url = {https://link.aps.org/doi/10.1103/PhysRevLett.131.053603}
}

@article{Schwartz2013,
  author  = {Schwartz, O. and Levitt, J. M. and Tenne, R. and Itzhakov, S. and Deutsch, Z. and Oron, D.},
  title   = {Superresolution microscopy with quantum emitters},
  journal = {Nano Letters},
  volume  = {13},
  number  = {12},
  pages   = {5832--5836},
  year    = {2013},
  doi     = {10.1021/nl402552m}
}

\end{document}